\documentclass[12pt]{article}
\usepackage{graphicx}
\usepackage{amssymb,amsmath,amsfonts,palatino,amsthm}
\usepackage{amssymb}
\usepackage{epstopdf}
\usepackage{slashed}
\usepackage{color}
\newcommand{\f}{\begin{equation}}
\newcommand{\ff}{\end{equation}}

\begin{document}

\title{Realism and causality I: \\
Pilot wave and retrocausal models as possible facilitators}
\author{Eliahu Cohen${}^{1}$, Marina Cort\^{e}s${}^{2,3}$,
Avshalom Elitzur${}^{4,5}$
and Lee Smolin${}^{3}$
\\
\\
${}^{1}$ Faculty of Engineering, Bar Ilan University, \\
Ramat Gan 5290002, Israel
\\
\\
${}^{2}$ Instituto de Astrof\'{i}sica e Ci\^{e}ncias do Espa\c{c}o${}^{3}$\\
Faculdade de Ci\^encias, Campo Grande, 1769-016 Lisboa, Portugal
\\
\\
${}^{3}$ Perimeter Institute for Theoretical Physics,
\\
31 Caroline Street North,
Waterloo, Ontario N2J 2Y5, Canada
\\
\\
${}^{4}$ Iyar, The Israeli Institute for Advanced Research, \\
POB 651, Zichron Ya'akov 3095303, Israel
\\
\\
${}^{5}$ Institute for Quantum Studies, Chapman University, \\
Orange, CA 92866, USA; }
\date{\today}
\maketitle
\newpage
\begin{abstract}

Of all basic principles of classical physics, realism should arguably be the last to be given up when seeking a better interpretation of quantum mechanics. We examine the de Broglie-Bohm pilot wave theory as a well developed example of a realistic  theory.  We present three challenges to a naive
reading of pilot-wave theory, each based on a system of several entangled particles.
With  the help  of a coarse  graining of pilot  wave  theory into a discrete  system,  we show  how these challenges can be answered. However this comes with a cost. In the description of individual systems, particles appear to scatter off empty branches of the  wave function as if they were particles, and conversely travel through particles as if they were waves. More generally, the ``particles'' of pilot wave theory
are led by the guidance equation to move in ways no classical particle would, involving
apparent violations of the principles of inertia and momentum conservation.


We next argue that  the aforementioned cost can be avoided within a retrocausal model. In the proposed version  of the pilot wave  theory,  the particle  is guided by a combination of advanced and  retarded waves. The resulting account for quantum physics seems to have greater heuristic power, it demands less damage to intuition, and moreover provides some general hints regarding spacetime and causality.

This is the first of two papers. In the second \cite{RETC} we show that, in the context of an explicit model, retrocausality, with respect to an effective, emergent spacetime metric, can coexist with a strict irreversibility of causal processes.

\end{abstract}
\newpage
\tableofcontents

\newpage

\section{Introduction: The Structure of this Paper}

This paper has an unusual structure. It is in the form of a dialogue, with challenges to the cogency of a realist formulation of quantum mechanics, specifically the de Broglie-Bohm (dBB) pilot wave theory \cite{dB1,Bohm1},  alternating with replies.  It is indeed the  record of an actual debate, carried out among the authors over two years.  It was a rare debate in that by the end
all four of us had significantly modified our positions, converging into a more coherent one which we now share.  Hence we felt it is interesting to try to preserve the structure of the debate in this paper.

The initial thesis of the debate was that the dBB pilot wave theory cannot
account for certain recently formulated thought experiments, without being so strained and  distorted, to the point of losing the claim to being  a realist  theory. These challenges took the form of three  proposed experiments, some of which  have  been  carried out, while the others are planned to be in the near term.



To answer these challenges, a  simplified version of dBB was developed, based on a coarse graining of the original pilot wave theory to a discrete configuration space, along the lines originally  introduced by Vink \cite{Vink}.  The assumption of discreteness serves as a main theme in the current series of papers, joining the primacy of causal relations and energy-momentum exchanges as our fundamental assumptions (see also the correspondence principles in \cite{FourPrinciples}). To formulate this coarse graining we articulated below four principles that one might ask of a realistic quantum theory.  We found that the fourth was inconsistent with the rest, and so had to be abandoned.  This encapsulates the
``cost'' of  achieving realism within dBB.  However, we show in detail that once this cost is paid, the coarse grained dBB theory, based on the other three principles, easily answers all three challenges.

The challengers next proposed a second thesis: Admitting retrocausality (and more generally, treating the initial and final states of the system on equal footing), might make possible a realist account of quantum physics at less cost to basic principles.
However, the existing retrocausal formulations of quantum theory, e.g. \cite{R1,R2,R3,R4,R5,R6,R6b,R7,R8,R9,R10}, are often expressed in an operational, instrumentalist language, in which ensembles are postselected as well as prepared (or preselected) \cite{PP1,PP2}. An additional difficulty is posed by the verification of a between-measurements state (actually a two-state \cite{TSVF}), which sounds like an oxymoron in quantum theory, and necessitates more delicate forms of measurement.

The synthesis of the debate is then a proposal to construct a retrocausal extension of dBB's
pilot wave theory (akin, but not identical to \cite{Rod1,Rod2,Sen}), that would avoid both the costs of dBB and the need to express retrocausal theories within an operational framework.  We do not propose a full-blown theory here, only the general structure and basic principles.

Finally, the debate took an unexpected turn when one of us  (MC) unexpectedly  found \cite{RETC} that a form of retrocausality occurs naturally in a model of dynamical causal structure they have been studying, called energetic causal sets \cite{ECS1}-\cite{DS}.  We propose to call this disordered causality.  Furthermore, while ECS have both classical \cite{ECS1}
and quantum \cite{ECS2} realizations, disordered causality is found already in the classical version.
This is discussed in a companion paper \cite{RETC}.

The structure of this paper then follows that of our debate: some brief opening remarks are followed by a series of challenges and answers.

\section{The Opening Challenge: Which Sacrifice Are We Willing to Make for a Better Theory?}

Every major revolution in physics has exerted its toll in the form of some major renunciation. Such was the case with geo-centrism banished by the Copernican revolution; perpetuum mobile outlawed by thermodynamics, and absolute space and time undermined by relativity theory. In all these cases, the notion sacrificed turned out to have been an obstacle for a better understanding of Nature. This, in fact, is one the hallmarks of scientific progress.

Quantum mechanics, however, has notoriously demanded a much greater price than its predecessors. Local realism was shown to be violated by quantum entanglement, and indeed at least one of three deeper notions has to be also compromised: Locality, determinism or the direction of causation. These have all been challenged by different schools, and even realism -- the very foundation of natural sciences -- has been dismissed by the Copenhagen interpretation, or rendered
irrelevant by currently popular operationalist formulations based on quantum information theory.  These revisions sharpen the language we use to describe quantum phenomena, while limiting its scope
to the description of manipulations we impose on quantum systems in the laboratory.

Which, then, may be the lesser evil? Bearing in mind the earlier revolutions, let the question be rephrased: Which sacrifice may turn out to yield the best advance in return?

In what follows we argue that abandoning realism is unlikely to make it up for physics in any significant way - after all, it has not done that so far.  At the same time, the existing realist approaches to quantum foundations, such as dBB or dynamical collapse models, entail issues which strain our expectations for a realistic description of nature. We discuss a series of  thought experiments that have been suggested as challenges to realism.  We analyze these in a version of dBB we call coarse grained pilot wave theory, in which the configurations space is coarse grained to a finite set of regions.
We find that dBB completely accounts for the phenomena, but it does so in ways that challenge our hopes and expectations for a realist description of the quantum world.

We seek to isolate the aspects of pilot wave theory that lead to this puzzling situation,
and we find out they include:

\begin{itemize}

\item{}The theory is causally asymmetric, in that the wave guides the particle, but the particle has no influence on the waves. This goes against Einstein's intuition as expressed e.g. in his formulation of the causal reciprocity between spacetime and mass.

\item{}The ``particle", or configurations do not obey the basic laws that we usually
take to define what we mean by a particle,
including the principle of inertia and the conservation of momentum\footnote{We should emphasize that this non-Newtonian behaviour of the dBB particle is known to at least some experts \cite{experts}, nor  would it have been a surprise to de Broglie himself, who noted ``The light quanta whose existence we assume do not always propagate in a straight line, as proved by the phenomena of diffraction. It then seems necessary to modify the principle of inertia.'' \cite{nosurprise}.}.

\end{itemize}

If dBB is the correct description of nature, would these issues be part of the price we have to
pay? Not necessarily. In what follows we explore an alternative which we find more natural. Relaxing
classical temporality, so as to allow quantum effects to proceed along both time directions, may open new vistas for the future.  As we explain in detail in the companion paper \cite{RETC}, we see this not as a renunciation of the hypothesis that there is a fundamental, irreversible time, whose activity is the continual creation of future events out of a (thick\footnote{That is the set of present
events may include some that are causally related; see \cite{ECS1}.}) present, but as an elaboration of that idea in which the arrow of time
defined by this continual creation of novel events, becomes disordered with respect to the time directions defined by the emergent Minkowski spacetime or the ticking of macroscopic clocks.

\subsection*{Does Bohmian Mechanics Alone Suffice?}

The dBB model is a realistic\footnote{Realist approaches to quantum foundations are the subject of recent and forthcoming books \cite{Becker,EUR}.} and  deterministic account that  assigns each  particle  ``hidden variables,'' in configuration space,  which,  although normally inaccessible  to measurement (in compliance with  the uncertainty principle), are presumed real.  It is based  on a simple  elaboration of the wave-particle duality, which  is that  both  waves  and  particles exist, the wave  guiding the particle.  When  describing single particles, dBB seems to capture most of our intuitive requirements from realism.

Difficulties arise, however, when  dBB addresses systems of entangled particles. Here,  the inherent nonlocality of quantum phenomena challenges our  intuitions,  because  in a hidden variable completion of quantum mechanics which  purports to describe the motion of every  individual system, the non-locality exposed by Bell's theorem must  be apparent in explicit detail.  Meanwhile, the mathematical abstraction  embodied by N-partite configuration space  seems  to strain  the common conceptions of realism. At some point,  the explanation given  by dBB becomes, to some, extremely baroque, and unconvincing.

Moreover, several  new Gedankenexperiments may appear to severely strain  dBB, forcing it to ascribe odder and odder properties to the initially  simple  wave-plus-particle description. In this paper we discuss three  examples of these challenges for realism, one based on two interacting Mach-Zehnder interferometers (MZIs), the others are variations of the three box puzzles.   They  are  highly  idealized, ignoring various technicalities, but  present situations in which the dBB account might be seen to be less appealing to some.

\section{The Basis for a Defence of dBB: A Coarse Grained Version of Pilot Wave Theory}

In this section we introduce a coarse graining of pilot wave theory,  along lines first given
in \cite{Vink},  which we will use to address the challenges to realism that will be presented
in this paper.

\subsection{Principles for a discrete coarse grained approximation to dBB}

In the exact theory of dBB the configuration space is a smooth manifold, and we posit that the evolution traces a continuous trajectory.
\f
x^a_I (t)   \in {\cal C},
\ff
which is the configuration space of $N$ particles, labeled by $I$ in a $d$
dimensional manifold.

A complete configuration consists of a wave function on $\cal C$ together with a position in $\cal C$.
\f
Z(t) = \{ \Psi (x,t)    , x^a (t) \}.
\ff
These evolve via the Schr\"odinger equation and the de Broglie guidance equation.


In the discussion of thought experiments it is helpful to be able to coarse grain
the configuration space into discrete regions.  This requires that we give up
on the continuous trajectories by which the dBB particles follow the guidance equation.
We must also give up on determinism. Instead we use probabilistic rules for
the particles to jump between regions.   This gives a form of pilot-wave theory
for discrete systems already explored by Vink in \cite{Vink}.

We start by formulating some principles that can guide us in the case that
the configuration space is discrete.  We will then use these to formulate a coarse graining of dBB which has a discrete configuration space.

The system is described by a configuration, which is an element of a discrete
configuration space, $x \in {\cal C}$ and a wave function $\Psi (x) $  on $\cal C$.
The configuration is  sometimes referred to as ``the particle".

We examine four  assumptions that one might want to assume for a coarse graining of dBB,
or any realist formulation of quantum theory, based on the above idea that there are both
waves and particles.

\begin{itemize}

\item{} {\bf (A)} The wave function evolves unitarily and independently of the particle's
configuration.  The particle evolves probabilistically, and the probabilities depend on the
wave function, through a coarse graining of the de Broglie guidance equation.

\item{} {\bf (B)}  The Born rule.  $\rho (x,t) = \psi^\ast(x,t) \psi (x,t) $
is the probability to find the particle at $x$ at time $t$.

In particular, the particle is never at a configuration $x_0$ at which $\psi (x_0 )=0$. \\

\item{} {\bf (C)}  The evolution respects that the particle is stationary when the
wave function is  real.

This is a consequence of the de Broglie guidance equation
\f
\dot{x}^a = \frac{1}{m} \nabla^a S,
\ff
where $S$ is the phase of the wave function
\f
\psi (x,t) = \sqrt{\rho}e^{\frac{\imath}{\hbar} S}.
\ff

Most importantly, we also want to impose a kind of locality, as part of a law of inertia.

\item{}{\bf (D)} Two particles continue in their states of motion or rest, so that momentum is conserved, except when they interact, and for that they have to coincide.

\end{itemize}

We will see below that {\bf (A)}, {\bf (B)}, {\bf (C) } and {\bf (D)} together lead to a contradiction.
In fact, while  {\bf (A)}, {\bf (B)} and {\bf (C) } are consequences of dBB,
{\bf (D)} is not, and in fact contradicts dBB, and the other three.  This is surprising at first,
but it is a known consequence of the de Broglie guidance equation.

\subsection{Construction of a coarse grained version of dBB}

We now construct a discrete coarse graining of dBB that respects the
first three principles, {\bf (A)}, {\bf (B)} and {\bf (C)}.

In this paper we work with a coarse graining of configuration space into a series of discrete
configurations, $ Z_t$, which evolve in a discrete time, $t=1,2, \dots ,  T$ where $T$ is the total
number of time steps.  The configuration space, at time $t$, $Z_t$, can also be time dependent.

We then have at each time an orthonormal basis $ | Z , t \rangle$ with $M_t$ elements.
A wave function is a normalized amplitude for each basis element,
\f
|\Psi  (t) \rangle = \sum_{i=1}^{M_t}  a_i (t) | i,t\rangle.
\ff
The Born rule probabilities are given by
\f
P_i (t) = |a_i (t)|^2.
\ff
We have discrete evolution between the time steps,
\f
|\Psi  (t+1) \rangle = \hat{U} (t+1, t ) |\Psi  (t) \rangle.
\ff

The dBB description is completed by an ensemble of particle positions.  We will not try to track
these individually but following \cite{Vink} we will just give rules to construct a probability distribution,
$\rho (i, t)$. Instead of a deterministic guidance equation we will then seek to give a transfer
function for the movement of particles among discrete states. Hence,
\f
T(j, t+1 ; i, t)
\ff
is defined to be the probability
that a particle in state $i$ at time $t$ will be in state $j$ at time $t+1$.   We have of course
\f
\rho (i, t+1) = \sum_j T(i, t+1 ; j, t) \rho (j, t).
\ff
Looking at
the simple examples in this paper, it is easy to see that we have coarse grained too much to  give a deterministic
description.

Finally, to account for the guidance rule and the Born rule, the transfer function between
time $t$ and $t+1$ must depend also
on the wave function at time $t$,
\f
T(j, t+1 ; i, t;  a_i (t) ).
\ff

\section{First Challenge to Realism: Two Particles Intersecting along their Mach-Zehnder Interferometers}

We are now ready to face the first case where this model is severely strained. We begin with the familiar MZI, which offers the simplest demonstration of the wave function's dynamics through its sensitivity to the relative phase. As long as no position measurement is made on the particle while traversing the MZI, it retains its initial momentum through interference. This makes it clear that the wave function has somehow traversed both MZI paths.

For Copenhagen, this is only natural. If nothing can indicate which path the particle has taken, then the path remains superposed just like the probabilistic distribution given by the equation. Any further assumption is consequently deemed superfluous.

\begin{figure}[h]
\centering \includegraphics[height=6cm]{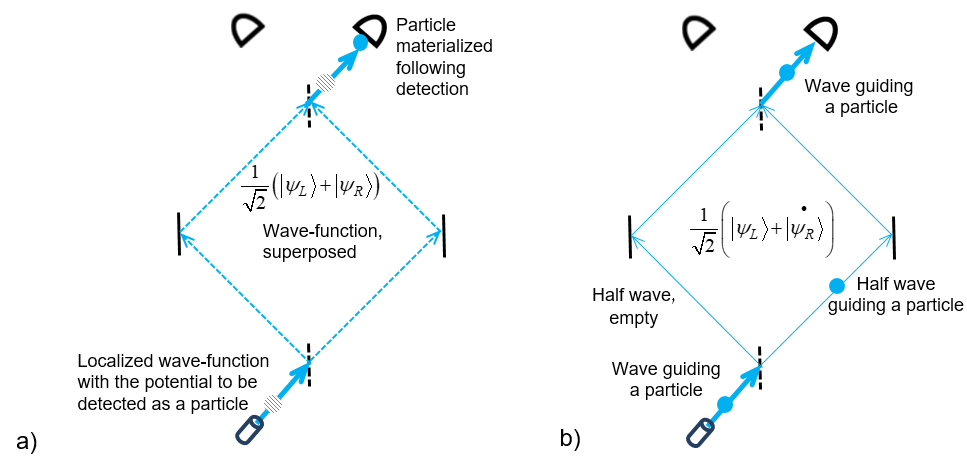}\caption{Single-particle interferometry according to the two interpretations. (a) Copenhagen Interpretation. (b) dBB }
\label{fig1}%
\end{figure}

The challenge to realism is to account for the role of the empty or ghost half of the wave function, within a realist account of the experiment.  For if the role of the wave function is to guide the particle, how can it matter what the value of the wave function is in a region of configuration space that is empty of the particle?

dBB's response to this challenge is ontological: The MZI was traversed by both a particle and its accompanying wave. The particle has taken one path together with half of the wave function while the other, empty half, took the second path. Because the two halves of the wave function may come together to interfere in the future,  the empty branch
potentially has a causal influence on the particle's motion. This is of course very elementary, but taking a minute to be clear here will help us think through more intricate examples to come. Fig. \ref{fig1} shows the two approaches to the MZI setting.


\subsection{We use the coarse grained version of dBB  to answer the first challenge}

To answer the first challenge,  we analyze the single MZI from the perspective of the coarse grained dBB. This serves as a warm-up exercise for more complex challenges to come.

We have three times, which we call $t_0,t_1,t_3$, to leave room for a stage in the middle when
we complicate the experiment.

\begin{itemize}

\item{$t_0 $}{ \bf - Before the particle enters the first beam splitter.}

There are two possible configurations, which can be taken to  label a momentum basis:
\f
C_0 =\{ + , - \}.
\ff
So the  two basis states for particles incoming to the first beam splitter are
\f
| + , t_0\rangle, \ \ \ |-, t_0\rangle.
\ff

\item{$t_1 $}{ \bf  - While the particle is between the first and the second beam splitters.}

After the first beam splitter the particle travels either to the right or the left, so
the possible configurations are now
\f
C_1 =\{ L, R \}.
\ff
So the  two basis states for particles between the two beam splitters are
\f
| L , t_1\rangle, \ \ \   | R, t_1\rangle.
\ff

\item{$t_3 $} {\bf - After the particle leaves the second beam splitter.}

The possible configurations are again
\f
C_3 =\{ + , - \}.
\ff
So the  two basis states for particles leaving the second beam splitter are
\f
| + , t_3 \rangle, \ \ \ |-, t_3 \rangle.
\ff

\end{itemize}

Next we discuss the wave function evolution at each evolution step.  At each step,
we write transition amplitudes, which define the time translation operator $\hat{U}$.
The single MZI has two evolution steps, at the first and second
beam splitters:

\begin{itemize}

\item{}{\bf The first beam splitter:  $t_0 \rightarrow t_1 $}

The first beam splitter has the following effect:
\f
| \pm , t_0 \rangle  \rightarrow | \pm , t_1 \rangle = \frac{1}{\sqrt{2}}  (  | L, t_1 \rangle \pm  | R, t_1 \rangle).
\ff

\item{}{\bf The second beam splitter:  $t_1 \rightarrow t_3 $}

The second beam splitter acts by time  reversal of the action of the first
\f
| \pm , t_1  \rangle  \rightarrow | \pm , t_3 \rangle
\ff
or, in the $L,R$ basis:
\f
| L , t_1  \rangle  \rightarrow \frac{1}{\sqrt{2}}  (  | +, t_3\rangle  +  | -, t_3 \rangle)  , \ \ \
| R , t_1  \rangle  \rightarrow \frac{1}{\sqrt{2}}  (  | +, t_3 \rangle  -  | -, t_3\rangle).
\ff
\end{itemize}

Having established the amplitudes for evolution of the wave function, we next prescribe
the probabilities by which the wave function guides the particles in the coarse grained dBB. These probabilities are specified for  transitions between configurations at consecutive times.
Given the guidance equation these depend on the wave function at a single time.

At the first beam splitter, we have:
\f
T( L, t_1 ; +, t_0;  | +,  t_0  \rangle)= T( R, t_1; + , t_0;  | +,  t_0  \rangle )=  \frac{1}{2}.
\label{Tt=0}
\ff
At the second beam splitter we have
\f
T( + , t_3 ; L, t_1; | +,   t_1  \rangle)=  T( + , t_3; R, t_1;  | +,   t_1  \rangle)=  1,
\label{Tt=1a}
\ff
but
\f
T( -, t_3; L, t_1;   | +,  t_1  \rangle)=  T(-, t_3; R, t_1;   | +,  t_1  \rangle )=  0.
\label{Tt=1b}
\ff

Note that these are required by agreement with the Born rule, condition {\bf (B)}.
But conservation of momentum, condition {\bf (D)}, is not respected, because it would
imply $L \rightarrow -$ at the second beam splitter.  But exactly because of the superposition principle, at the second beam splitter the $|+ , t_1\rangle$ state goes all $|+\rangle$.  Therefore
those particles coming along the L beam have to swerve around to $+$ at the second
beam splitter, rather than continue ahead as condition {\bf (D)} would require. Hence, conditions {\bf (A)} and {\bf (D)} are in conflict and, if we assume that the guidance rule is designed to
preserve the Born rule, the conservation of momentum is sacrificed.

\section{Second Challenge to Realism: Two Intersecting Mach-Zehnder Interferometers} \label{crossing}

\subsection{Introducing Incomplete Measurements}

Interaction-Free Measurement (IFM) has further stressed the wave function's peculiar nature by showing that even the {\it non-clicking} of a single detector along one of the MZI paths disturbs the interference \cite{IFM}. A series of Gedankenexperiments then followed \cite{HardyP,Hardy,Partial,Nonseq,ED,Oblivion,TooLate,Liar}, almost as simple yet revealing even more paradoxical consequences of the formalism. Among the wide class of ``incomplete measurements'' \cite{EC}, these are classified as ``delayed measurements.'' They involve two or more particles ``measuring'' one another only by entanglement, before the macroscopic detectors complete the measurement process. Thus Hardy \cite{Hardy} has demonstrated entanglement of two distant particles by virtue of their simultaneous interaction with the two distant halves of the wave function of a single particle located between them. He has also produced the Hardy Paradox \cite{HardyP} where a particle and anti-particle interact without annihilating one another. Elitzur and Dolev continued this line of research with a Gedankenexperiment demonstrating the wave function's non-sequential dynamics \cite{Nonseq}, and with the quantum liar paradox \cite{ED}, recently presented in a more straightforward form with a few variants \cite{Liar}.

Some of these experiments were analyzed along the lines of dBB \cite{dBBA1,dBBA2}, as well as retrocausal models \cite{retroA1,retroA2,retroA3,retroA4,retroA5}; see \cite{Rod1,Rod2,Sen,TooLate,Grif,Hartle,Wiseman,Stein1,Stein2}, calling for comparison between these approaches, which, as stated earlier, we believe can be naturally integrated together. DBB had to stress that even the empty part of the wave function can exert causal effects on the other particle. The retrocausal models, invoking the additional wave function coming from the future, argued that, during the time interval between two quantum measurements, a particle must manifest phenomena even more alien to classical intuitions than hitherto believed. Weak measurement \cite{AAV}, the experimental offshoot of the Two-State-Vector Formalism (TSVF), was pointed out as the sufficiently delicate technique to reveal such special situations \cite{PP1,AAV}, and recently even projective measurements were shown to be effective for this purpose \cite{Shutter1,Shutter2,Disappearing,Spooky,CP}.

Below we present a simple variant of Hardy's paradox, which utilizes strong (projective) quantum measurements rather than weak measurements.  We then discuss the extent to which this example  can strain Bohmian mechanics and reveal the possible advantage proposed by retrocausal approaches.

\subsection{The experiment:  Can an empty wave proceed after absorption? }

An important feature of the Bohmian ``guide wave'' is that, like a classical wave, it must be obstructed (absorbed or reflected) by any obstacle which could interact with its associated particle. This is in fact obliged by the model's adherence to realism. Therefore, the empty part of the wave thus disturbed along its path in the MZI does not take part in the consecutive interference, thereby spoiling it. It is this feature that, in the following simple experiment, leads to an apparent conceptual difficulty.

\begin{figure}[h]
\centering \includegraphics[height=5cm]{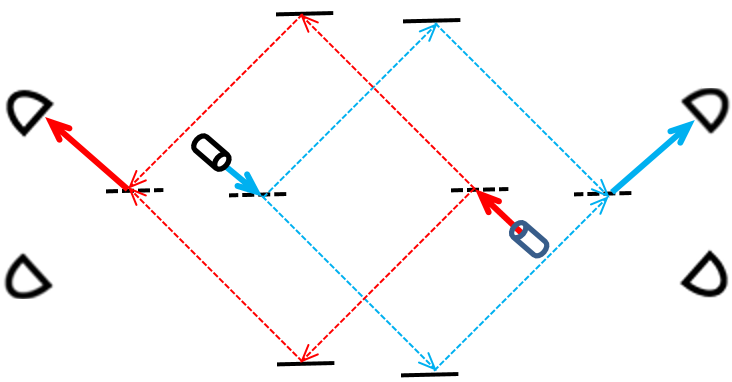}\caption{The proposed experiment. When no annihilation occurs, both positions and momenta become correlated.}%
\label{fig2}%
\end{figure}

Let two MZIs be placed such that their paths symmetrically cross each other at two points (Fig. \ref{fig2}). Let one MZI measure an electron while the other is traversed by a large, positively charged molecule. The molecule's size is such that, has its path simply crossed that of the electron while each of them is fairly localized, the molecule would always absorb the electron. The combined neutral molecule would then be deflected out of the interferometer.

Suppose next that (taking care of perfect timing for the two wave functions to meet at the crossing point), this absorption did not take place, that is, both the electron  and the molecule emerge from their MZIs, the latter remaining positively charged. The two particles are now entangled:
\begin{equation}
\frac{1}{\sqrt{2}}\left[|\psi_R^1\rangle|\psi_L^2\rangle+|\psi_L^1\rangle|\psi_R^2\rangle\right],
\end{equation}
which means that, if we measure their positions within their respective MZIs, they will be strictly correlated, R-L (being the right/left arms). Alternatively, if we wait for their interference effects, these will manifest the same $+-$ correlation (where $+$ and $-$ correspond to each of the two output ports of the MZIs).

These correlations are nonlocal in the strictest EPR-Bell sense.
Suppose then, that in the two MZIs, Stern-Gerlach magnets serve as beam splitters,
splitting the electron and the molecule according to their spins along some direction. Should non-absorption occur, they are now correlated along any spin direction. As the different spin directions maintain the same uncertainty relations as position and momentum, the ubiquitous EPR-Bell correlation oblige that the electron's/molecule's wave function must travel through both MZIs paths.

A challenge  to a fully realistic picture is now visible. i) By the uncertainty relations, each spin direction is maintained through the interference of the wave function's two components. But ii) The above setting guarantees that, even when the electron's and the molecule's corpuscles do not collide (which would end up in the electron's absorption by the corpuscle, which we discard), each corpuscle obstructs the other particle's empty half wave. {\it It would seem that no interference can thus take place, hence neither EPR-like correlations. } Yet iii) Correlations for all spin directions are still obliged by QM. In other words, it seems that the molecule's/electron's presumed empty half wave is not obstructed by the electron's/molecule's corpuscle, going through it as if it was not there!

\begin{equation}
\frac{1}{\sqrt{2}}\left[\dot{|\psi_R^1\rangle}\dot{|\psi_L^2\rangle}+|\psi_L^1\rangle|\psi_R^2\rangle\right]\rightarrow\dot{|\psi_R^1\rangle}\dot{|\psi_L^2\rangle}.
\end{equation}

\begin{figure}[h]
\centering \includegraphics[height=5cm]{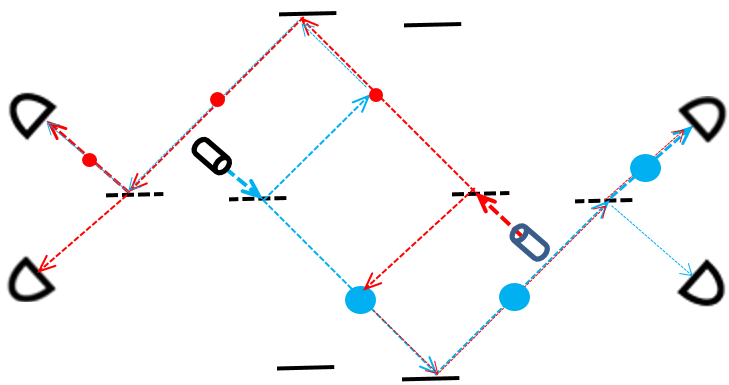}\caption{The putative Bohmian explanation for the predicted effect. The feeble dashed lines represent dBB empty waves absorbed by the corpuscles.}%
\label{fig3}%
\end{figure}

\subsection{Coarse grained dBB responds}

How can Bohmian mechanics resolve this seeming difficulty while maintaining some degree of mechanistic realism we look for?  A possible answer is evident if we go through the description of the experiment in detail.  The key is principle {\bf (A)}, according to which the wave guides the particle, but the particle has no effect on the wave.  The problem emerges from the fact that not even the presence or absence of the particle can affect the wave.  Consequently the wave evolves on the configuration space independent of where the particle is, so the same projection down to the entangled state occurs for the components of the wave function that are not directed out of the interferometer in the cases
where the electron and molecule collide and combine.  That is, because the wave function is in
a superposition of branches that do intersect at the crossing points, and branches which do not, the former can be projected out of the wave function, leaving and entangling the latter.  By {\bf (A)}
this goes on whichever the particles happen to be, and there are members of the ensemble in which the particles take every possible route.

\begin{figure}[h]
\centering \includegraphics[height=8cm]{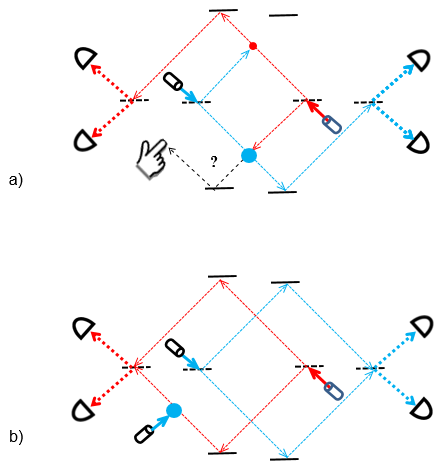}\caption{A modification examplifying a problem with the Bohmian explanation. a)  Why should the finger inserted in the apparently-empty path spoil the momenta correlation? b) Indeed, the same spoiling occurs when an additional (but non-superposed) molecule is placed on the electron's MZI and turns out not to have blocked it.}%
\label{fig4}%
\end{figure}

How can a realistic picture account for this oddity? One might try to think of it as follows. These half-waves, each blocked (absorbed or scattered) by the other party's corpuscle, keep guiding the corpuscle of their ``absorber.'' Even worse, the problem with this suggestion is that it makes the evolution of the wave function depend on where the particles are, which is forbidden in dBB.  Hence, this violates
{\bf (A)} because it implies that the wave is influenced by whether or not the particles (corpuscles) collide. In any case, this explanation is ruled out by the following modification: Place any obstacle behind the crossing point, where the molecule's/electron's empty wave is supposed to have been blocked by the electron's/molecule's corpuscle, respectively, and the correlations would vanish (Fig. \ref{fig4}a).

To bring  the significance of this effect closer to home, notice that the obstacle can be just another molecule/electron identical to those  traversing the MZI, except  that this time it is not superposed but resides on that  path only (Fig. \ref{fig4}b).  This particle is of course localized, but for a truly realistic dBB framework, it is no more localized than the superposed molecule/electron whose corpuscle happened to reside on this path. So if this localized particle suffices to ruin the interference, how can the superposed one allow the other party's half wave to proceed undisturbed?

To clarify how dBB answers this challenge we next  go through how dBB describes the
double MZI experiment. We describe the double MZI experiment in the coarse grained dBB, where now we insert
a fourth time, $t_2$ between times $t_1$ and $t_3$.

\begin{itemize}

\item{$t_0 $}{ \bf Before the particles enter the first beam splitters.}

There are four possible configurations, two for each particle
\f
C_0 =\{ ++, +-, -+ , -- \}.
\ff
So the  four basis states for particles incoming to the first beam splitter are
\f
| \pm , t_0\rangle \otimes | \pm , t_0\rangle =  | \pm , \pm , t_0\rangle.
\ff

\item{$t_1 $}{ \bf  While the particles are between the first beam splitters and the crossing points.}
Again, we double the configuration space and the corresponding configuration basis.
\f
C_1 =\{ Ll, Lr, Rl, Rr \}.
\ff
So the four basis states for particles between the two beam splitters are
\f
| L l , t_1\rangle \otimes | l l , t_1\rangle =| L l , t_1\rangle, \dots
\ff

\item{$t_2  $} {\bf After the particles leave the crossing points.}  There are now two more
possible outcomes.  If the particles meet at the $R,r$ crossing point they are deflected
upwards into new configurations which we may call $T,t$.  If they meet
at the $L,l$ crossing point they are deflected to $B,b$.

The possible configurations are
\f
C_2 =\{ Ll, Lr, Rl, Rr, Tt, Bb \},
\ff
although we note that $L,l$ and $R,r$ are never occupied.
Crossing points are
\f
| Lr , t_2 \rangle, \ \ \ | Rl , t_2 \rangle,| Tt , t_2 \rangle,| Bb , t_2 \rangle.
\ff

\item{$t_3  $} {\bf After the particles leave the second beam splitter, or are diverted.}

The possible configurations are the direct product of those for each particle, plus the two diverted pathways. However, as a last step we postselect on there being nothing in either of the diverted
pathways so we have
\f
C_3 =\{ ++, +-,-+, --,Tt, Bb \} \rightarrow \{ ++, +-,-+, -- \},
\ff
and each of these label a basis state.
\begin{eqnarray}
& | ++ , t_3 \rangle, &  | +- , t_3 \rangle| -+ , t_3 \rangle,| -- , t_3 \rangle,| Tt , t_3 \rangle,| Bb , t_3 \rangle,
\nonumber
\\
& \rightarrow & | ++ , t_3 \rangle, | +- , t_3 \rangle| -+ , t_3 \rangle,| -- , t_3 \rangle.
\end{eqnarray}

\end{itemize}

We now have three evolution steps.  As in the warm up, we first  discuss the wave function evolution at each step.
We give transition amplitudes, which define the time translation operator $\hat{U}$.

\begin{itemize}

\item{}{\bf The first beam splitter:  $t_0 \rightarrow t_1 $}

The first beam splitter has the following effect:
\f
| \pm , t_0 \rangle  \rightarrow | \pm , t_1 \rangle = \frac{1}{\sqrt{2}}  (  | L, t_1 \rangle \pm  | R, t_1 \rangle).
\ff

\item{}{\bf The second beam splitter:  $t_1 \rightarrow t_2 $}

The second beam splitter acts by time  reversal of the action of the first
\f
| \pm , t_1  \rangle  \rightarrow | \pm , t_3 \rangle
\ff
or, in the $L,R$ basis:
\f
| L , t_1  \rangle  \rightarrow \frac{1}{\sqrt{2}}  (  | +, t_3\rangle  +  | -, t_3 \rangle)  ,
| R , t_1  \rangle  \rightarrow \frac{1}{\sqrt{2}}  (  | +, t_3 \rangle  -  | -, t_3\rangle).
\ff
\end{itemize}


The evolution rules at the two beam splitters are just the product of those given in the
single interferometer.  At the crossing point, $t_1 \rightarrow t_2$ we have the evolution rules
\begin{eqnarray}
|R\rangle \otimes |r\rangle & \rightarrow & |T\rangle \otimes |t\rangle
\nonumber
\\
|R\rangle \otimes |l\rangle & \rightarrow & |R\rangle \otimes |l\rangle
\nonumber
\\
|L\rangle \otimes |r\rangle & \rightarrow & |L\rangle \otimes |r\rangle.
\end{eqnarray}

We can then easily compute the evolution of the wave function
\begin{eqnarray}
&t_0&:    |\Psi (t_0)\rangle = |+\rangle \otimes |+\rangle
\nonumber
\\
&t_1&:    |\Psi (t_1)\rangle = |+\rangle \otimes |+\rangle= \frac{1}{2} ( |L\rangle \oplus  | R \rangle )\otimes (|l\rangle \oplus |r \rangle)
\nonumber
\\
&t_2&:    |\Psi (t_2)\rangle = \frac{1}{2} ( |T\rangle  |t\rangle \oplus  ( |B\rangle | b \rangle )
\oplus  \frac{1}{2} (|L\rangle   |r \rangle \oplus |R\rangle  |l\rangle)
\nonumber
\\
&t_3&:    |\Psi (t_3)\rangle =  \frac{1}{2} (|+\rangle   |- \rangle \oplus |-\rangle  |+\rangle).
\end{eqnarray}




Finally we give the transfer functions for probabilities  for the particle configurations,
in the coarse grained dBB.
At the beam splitters these are products of the individual probabilities we
have already specified in (\ref{Tt=0},\ref{Tt=1a},\ref{Tt=1b}). We just need to determine
the transition probabilities at the collision points

\f
T(Rr \rightarrow Tt  | +,  t_2  \rangle)= T(Ll \rightarrow Bb  | +,  t_2  \rangle )= 1
\label{Tt=2a}
\ff
\f
T(Lr \rightarrow Lr  | +,  t_2  \rangle)= T(Rl \rightarrow Rl  | +,  t_2  \rangle )= 1
\label{Tt=b}
\ff
with the rest being zero.

We can now write the evolution of the probability distribution for the configurations.

\begin{eqnarray}
& t_0 &:  100\% (+,+)
\nonumber
\\
& t_1 &:  25 \% \left [ (L,l)    + (L,r)     + (R,l)   + (R,r)      \right ]
\nonumber
\\
& t_2 &: 25 \% \left [ (B,b)    + (T,t)     + (R,l)   + (R,r)      \right ]
\nonumber
\\
& t_3 &: 25 \% \left [ (B,b)    + (T,t)     + (+,-)   + (-,+)      \right ]
\end{eqnarray}
Postselecting out the outcomes $(B,b)$ and $(T,t)$ we are left with
$50 \%$ of the runs, distributed as
\f
 25 \% \left [ (+,-)   + (-,+)      \right ]
\ff

We thus successfully reproduce the correct probabilities for the double MZI experiment with two crossed
MZIs, including the generation of the entangled output after postselection.
We see that the cost is what   we paid before to get the single MZI to work:
an asymmetrical, non-reciprocated action of the wave on the particle, and the breakdown
of conservation of momentum and the principle of inertia.

\section{Third Challenge: The Three Boxes Experiment in the ``Shutter Version''} \label{shutter}

Another challenge to dBB is posed by a recent intriguing prediction of TSVF \cite{Shutter1,Shutter2}.

\subsection{The experiment}

Consider an atom which goes from an initial state $|Init\rangle$ to one of three boxes.
The states corresponding to its position within the boxes are denoted by $|A\rangle $, $|B\rangle $, and $|C\rangle $.
At time $t_1$ we use a beam splitter to distribute the atom in a superposition of the three box states.
\f
|Init \rangle \rightarrow \frac{1}{\sqrt{3}} \left( |A\rangle  + |B\rangle + |C\rangle  \right).
\ff
The location of the atom is to be determined by passing a photon through boxes $A$
and $B$.   The photon in state $|a \rangle$ passes through box $A$, etc.
To this end we prepare a photon in an initial state $| \gamma \rangle$ and then, also
at $t_1$ we use a
photon beam splitter to split it into two states $|a \rangle$ and $|b\rangle$
\f
|\gamma \rangle \rightarrow \frac{1}{\sqrt{2}} \left( |a\rangle  + |b\rangle  \right).
\ff
We pass the photons through their respective boxes at time $t_2$.
If the atom is in box  $A$, the photon in state $a$ will be reflected from the box
\f
|A\rangle |a\rangle \rightarrow |A\rangle |a R\rangle.
\ff
If the atom is in one of the other two boxes the photon is transmitted through the box
\f
|B,C\rangle |a\rangle \rightarrow |B,C\rangle |a T\rangle,
\ff
with the analogous rules for a photon at $b$.

Thus the evolution proceeds as
\begin{eqnarray}
|Init\rangle|\gamma \rangle & \rightarrow & \frac{1}{\sqrt{6}} \left( |A\rangle  + |B\rangle + |C\rangle  \right) \left( |a\rangle  + |b\rangle  \right)
 \\
 & \rightarrow & \frac{1}{\sqrt{6}} \left( |A\rangle|aR\rangle + |A\rangle|bT\rangle  + |B\rangle|aT\rangle +|B\rangle|bR\rangle + |C\rangle |aT\rangle  + |C\rangle |bT\rangle  \right)
 \nonumber \\
 &=& |Output\rangle.
\end{eqnarray}

We next, at $t_3$ pass the atom through a filter which projects out the component in
the state
\f
| F\rangle = \frac{1}{\sqrt{3}} \left( |A\rangle  + |B\rangle -  |C\rangle  \right).
\ff
This gives us a final state for the photon
\f
|\gamma , final \rangle = \langle F| Output \rangle = \frac{1}{9}\frac{1}{\sqrt{2}}
\left( |aR\rangle  + |bR\rangle  \right).
\ff

We note that the projection on $|F\rangle$ yields a  final state only $\frac{1}{9}$ of the time we run the experiment.  We ignore the rest of the runs and study the results of the projection
(i.e. we treat the projection as a postselection.)
But focusing on those $\frac{1}{9}$ of the runs, it seems that the photon is {\it reflected from both} $a$ and
$b$. It is {\it never transmitted}, and moreover, the photon maintains its coherent superposition in this fraction of cases. This logic will be better understood in the following:

We are free at any later time $t_4$ to ask if the atom was in Box A.  We do this by measuring the photon in the $a$ channel to see if it is in the transmitted channel or the reflected channel.

In the ensemble defined by the selection or preparation on $| Init \rangle$ and the projection
or postselection on $|F\rangle$ we
always find the photon in the reflected channel of Box A, so in these cases the answer is yes, the atom was in Box A.

But we could have asked instead the same question about box B, by checking for  a photon in the  $b$ channels.  The answer would have been the same: the photon is always in the reflected channel.  Thus it would seem that with the preparation and projection as defined, the atom is in whichever box we look. Moreover the choice as to which box we look for the atom in can be made long after the photon passed through the boxes, by a choice of which photon channel to query.

Note that there is only one photon, so  we can only look for it in the $a$ channel or $b$ channel, not both. But whichever we choose, in the ensemble defined in this experiment, we will always find the photon has been reflected by the atom in the channel we choose to look for it.  It is as if the atom were in both boxes. This is evident when performing an interference experiment of the reflected photon. In the cases on which we focus, the photon would constructively interfere with itself, suggesting it was never transmitted through either box.

\subsection{The dBB response}

How would dBB describe this experiment?
To answer, we set up the same kind of coarse grained dBB model we have previously defined.


Pilot wave theory tells us to add a particle to the wave associated with the atom, and one to the wave associated with the photon.  Each is conserved, so at any instance there is one atom particle and one
photon particle.  The new degree of freedom is hence a point in the joint
configuration space of the atom and photon.
The challenge is that this configuration space is discrete and changes with time.

As we are about to see, it is indeed not obvious how  to invent a version of Bohmian mechanics that is adequate to the situation.  It must address what happens to the particle degree of freedom under interactions between particles, projective measurements etc.; when the configuration space is discrete and when it changes as the experiment proceeds.

The configuration space is time dependent.  At $t_1$ the configuration is
\f
X (t_1) = (I,i), \ \ \ \ I=A,B,C; \ \ \ \ i=a,b.
\ff
At $t_2$ we add the information as to whether the photon is reflected or transmitted
\f
X (t_2) = (I,i,W), \ \ \ \ I=A,B,C; \ \ \ \ i=a,b \ \ \ \ W=T,R.
\ff

The wave function is real at each time so the current is zero and the particle stays where it is;
hence $I$ and $i$ are conserved. There is one evolution step when the photon passes through a box. The question is how to handle it.

Principle {\bf (D)} requires that the atom particle and photon particle can only interact and hence alter their states when they are in the same box.
This implies a simple rule, which represents the local character of the interaction between the atom and the photon.
\f
W=R \ \ \mbox{if} \ \  I = i  \ \ \ \ \  W=T \ \ \mbox{if} \ \ I \neq  i.
\ff

At $t_3$ the configuration space is reduced to the four possible configurations of the
photon
\f
X (t_3) = (i,W),  \ \ \ \ i=a,b \ \ \ \ W=T,R
\ff
with the projection from $t_2$ to $t_3$ preserving the values from $t_2$.   That is
\f
(i,W)_{t_3} = (i,W)_{t_2}.
\ff

Pilot wave theory treats probability as a property of an ensemble defined by
running  the experiment with the same initial wave function but with different values
of the initial configuration variable distributed according to Born's rule.

We carry out a large number of runs of the experiment, which produces an ensemble,
defining a probability distribution $P(X,t)$. By using the Born rule
\f
P(X,t) = \rho (X, t),
\ff
we know that initially the ensemble consists of an equal distribution of the particle in boxes $A,B$ and $C$.  In each of these cases the photon is half the time initially  in the
state $a$ and half the time in the state $b$.

We note that the wave function is always real.  Hence $I$ and $i$ are conserved and the atom and photons stay where they are initially for the full run of the experiment.

Let us consider the $\frac{1}{3}$ of cases in which the particle corresponding to the atom is initially at $I=C$.  In half these cases the photon is at $a$ and in the other half the photon is at $b$.

Do any of the cases with $I=C$ make it into the ensemble which is the result of
projecting on $|F\rangle$?  Since we have assumed the validity of Born's rule we can
ask what is the probability that $|F\rangle$ as $|C\rangle$.  This is
\f
\left | \langle F| C \rangle \right |^2 =\frac{1}{3}.
\ff
So $\frac{1}{3}$ of the experimental runs that have a state after the projection had
$I=C$ initially.

Because the photon particle is never in the box with the atom particle in $100 \%$ of the
cases after time $t_2$ the $W$ value will be $W=T$.  This is a consequence of locality,
which  is assumption {\bf (D)}.

But this contradicts {\bf (B)} because after the projective measurement at $t_3$
\f
\rho_{t_3} (i, W=T ) =0.
\ff

Hence we have to give up at least one of {\bf (A)}, {\bf (B)} or {\bf (C)}.

If we drop the locality assumption we can posit that there is no evolution rule, instead the
particles just distribute themselves so that Born's rule is satisfied.

Hence, at $t_3$ the photon particles in the ensemble distribute themselves
in the two states $(i,W) = (a, R)$ and $(b,R)$.
This is in accord with {\bf (B)}, but it means that in the cases where $I=C$, in which the atom's particle starts at $C$ and stays at $C$, the photon reflected off a particle that was not in the box with them. The latter counter-intuitive interaction is exactly the one which seems to put an obstacle in the way towards realism.

Of course we already know that dBB for composite systems is highly nonlocal, so this does not rule it out, but this shows how weird you have to get to make it work. Taking seriously that the wave guides the particle implies non-local scattering where particles (and even empty waves!) reflect off of empty boxes.

In summary, it looks like such cases require further revision of dBB such that the empty wave is endowed with corpuscular properties, and vice versa, the corpuscle should sometimes act like a wave.

\section{A Dynamical Three Boxes a-la Bohm -- and a New Paradox}

Recently, an even more striking variant of the above paradox was introduced \cite{Disappearing,Spooky}. Assume that Boxes A and B are connected so that a particle can tunnel from one to the other and back, but the remote Box C remains unconnected to the other two. Under a special combination of pre- and postselected states, then, for a specific choice of 3 intermediate times, the following predictions hold:
\begin{itemize}
\item{} Has one opened Box A at time $t_1$, one would find there the particle with certainty.

\item{} Has one opened Boxes A and B at time $t_2$ they would both be empty.

\item{} Has one opened Box B at time $t_3$, one would find there the particle with certainty.
\end{itemize}

Using the pre- and postselected particle as a shutter, we can send towards it a probe particle superposed in space and time such that it would arrive to Box A at time $t_1$, to Box C at $t_2$ and to B at $t_3$. The probe particle is expected to return from all these different locations and instances in a coherent superposition, indicating the nonlocal disappearance and reappearance of the shutter particle in all these boxes.\\

How could dBB explain these predictions? Similarly to the above explanation of the simpler paradox, it seems that dBB would have to grant waves with particle properties and vice versa, but now in a time-dependent way. We leave it as a problem for the reader to show how the coarse grained version of dBB can be set up in this more complicated experiment and does indeed resolve the problem. In the next section we consider an alternative account. \\

\section{Towards a Retrocausal Realistic Formulation}

Aspiring to preserve realism, the dBB model invokes both a particle and an accompanying guiding wave. Once, however, we apply this model to some simple interactions between particles, the model seems to become cumbersome to the point that some might wonder whether the medicine is not worse than the disease, namely, Copenhagen and the abandonment of realism. We believe, however, that a certain twist can make dBB simpler and more elegant. The idea is simple: {\it What if half of the wave-like properties comes from the future and half comes from the past?}

\subsection{How retrocausal theories might answer the challenges}

\label{section:quantumretrocasual}

We suggest that retrocausal models might offer a straightforward way to overcome all the difficulties we have been discussing (see also a previous, and in many senses complementary approach \cite{Rod1,Rod2}). At first this is not so evident, because the existing retrocausal formulations of quantum mechanics are presented in an operationalist setting. We want to propose instead a retrocausal modification of pilot wave theory, which retains its realist approach to physics.

What we learn from the existing retrocausal formulations of quantum theory is that for a quantum interaction between two subsystems to reveal its full significance,  postselection (namely the final measurement that picks the cases that did not end up with absorption) is as causally essential as is pre-selection (namely the initial preparation with the first beam splitter). Consider again the basic double-slit experiment. Initially, there is a forward-evolving wave function (Cramer's ``offer wave'' or Aharonov's ``history vector''). Similarly to a classical wave, it traverses both paths, reaching the second BS from both sides and then proceeding to the two detectors. It is the reciprocal wave function returning from these detectors (Cramer's ``confirmation waves'' or Aharonov's ``destiny vector'') that determines the corpuscle's final position. Notice that the combination of the two evolutions along both time directions creates not only the full corpuscular trajectory but also the other, ``empty'' trajectory, which seems to have been traversed by ``nothing.'' However, this ``nothing'' was claimed to be the result of destructive interference in the transactional interpretation \cite{R4,R5}, while being a result of particle and ``negaparticle'' within the TSVF \cite{Anomalous,WR}. In such time-symmetric approaches the interplay between past and future boundary conditions may thus alleviate spatial peculiarities \cite{Future}.

\subsection{What would a realist retrocausal theory look like?}

Admittedly, the introduction of retrocausality is not a minor revision in the general picture of physics because it takes us to the much wider issues concerning the nature of time. The mainstream relativistic ``block universe'' treats time as the fourth dimension alongside with the three spatial ones, rendering time's apparent passage a subjective illusion. Rather, all events -- past, present and future -- are considered equally real like different locations in space. Several problems associated with this counter intuitive picture have long ago led to an alternative account, still open ended, namely ``becoming'' \cite{Becoming1,Becoming2,Becoming3,Becoming4,Becoming5}, where time is taken as much more profound, and more akin to its naive image. We can call this an {\it ``active"} notion of time.  According to it, there is a fundamental distinction between the past, present and future. In some sense, time is the very coming into being of events one after another, as the ``now'' proceeds from past to future \cite{Becoming2}. Trajectories, evolutions and histories genuinely ``grow'' into the yet-non-existent future. The past, on the other hand, is where the current block universe model can be regarded as perfectly valid, i.e. events are fixed and unchangeable, obeying well defined causal relations.

This is one foundation of our proposed approach to quantum mechanics. The second foundation is a certain form of time-symmetry. Within the TSVF, a measurement's effects proceed to both time directions, namely towards both future and past, until the next/previous measurement.

These two foundations would seem  to contradict each other.  Our  main  message here  is that  they  need not. The reason  is that  we have  to separate our two distinct aspects of time.  The first is the fundamental causal  activity  of time. The second  is the embedding of that  causal  process  within an emergent, coarse  grained description.

The  first,  causal  aspect of time -- the active one -- never reverses.  An  event,  once  happening, cannot  be made  to ``unhappen.'' Even if an event  is followed by a second  event which  undoes its action -- this is still a sequence of two  events.   But what can reverse is the direction of causal  arrows in the embedding of the fundamental causal  order  in the emergent spacetime.  This  is the  operation we  refer  to when  we  speak  of microscopic time-symmetry.

In the companion paper \cite{RETC} we explicitly  show, in a classical  model,  these  two aspects of time -- the active  irreversible, creative  (in Bergson's  sense)  one,  and  the emergent symmetric spacetime.

Time-symmetric approaches for quantum mechanics have been investigated in the last few decades from various perspectives, starting from \cite{R1,R2,R3}, to the development of the Transactional Interpretation \cite{R4,R5}, the Two-State-Vector Formalism (TSVF) \cite{PP1,PP2,TSVF} and its more recent extension, the Two-Time Interpretation \cite{TTI1,TTI2,TTI3}, as well as other general considerations \cite{R6,R6b,R7,R8,R9,R10}. These approaches have their differences, but they share the motivation of retrieving microscopic time-symmetry which seems to be lost upon collapse.

The following is a simple example. Consider a photon emitted by a source S towards a beam splitter behind which are two detectors A and B. Let A be much closer to the beam splitter than B. Suppose then that A, to which the wave function arrives first, does {\it not} click. We immediately know that the photon is surely on the other path going to B. How did this change occur? In the framework of our retrocausal model, a backward-evolving wave emitted from A cancels the right-hand wave function's half, by destructive interference, all the way back to the source. Then, via the spacetime zigzag, strengthens, by constructive interference, the other half going to B and completes it to one. This is why, upon A's mere silence, the wave function ``collapses'' so as to make the photon certainly reside on the path leading to B. Once there is one full wave from S to B, a particle trajectory is formed. Similarly for the reverse case where the detection occurs at A and cancels the possibility of detection in B: The wave emitted from A completes the transaction while the source now sends a cancelling wave to B. This holds even for the cases where A and B receive their halves together, although the present relativistic framework does not allow saying (or even asking) which detector has reacted ``first'' in the ordinary temporal order.

With these two foundations we aspire to revise dBB to a simple and natural interpretation. What are the pros and cons for this model? Let us begin with the latter, which are admittedly obvious. We are talking about {\it evolution} in spacetime as if there is another time parameter, about which we yet know nothing. But this is not necessarily a disadvantage for a foundational physicist who might be long suspecting that something about time is still missing in the ``block-universe'' view.

The pros for introducting becoming, on the other hand, are also clear, and moreover greater. Allowing the forward- and backward-evolving waves to interact along time with constructive and destructive interference is an appealingly simple idea first introduced by Wheeler and Feynman \cite{WF} to account for classical electromagnetism, later applied to gravitation and cosmology by Hoyle and Narlikar \cite{HN1,HN2}. Cramer then applied these ideas to quantum mechanics \cite{R4}. This brings a great deal of simplicity to the quantum world.

Take for example the EPR experiment: What appears to be a nonlocal influence between two measurements faraway in space becomes perfectly local in spacetime through the so-called Parisian Zigzag \cite{Zigzag}, allowing remote events to affect each other retrocausally by taking advantage of their common past. Similarly for all varieties of quantum oblivion \cite{Oblivion,Oblivion2,Oblivion3}, where an event that merely {\it could} have occurred leaves a physical effect even when it {\it did not}. In a simple interaction between a particle and anti-particle as shown in \cite{Oblivion,Oblivion2,Oblivion3} it has been proved that the two particles went through a brief period of entanglement, followed by a mutual cancellation of this entanglement, leaving one of them localized, while the other remains unaffected. Retrocausality easily resolves the paradox. Even more so, the retrocausal account easily handles other famous temporal quantum oddities like Wheeler's \cite{Wheeler} delayed-choice experiment or the ``quantum liar'' paradox \cite{ED,Liar}.

Another significant advance converging into this direction is Aharonov's Two-State Vector Formalism (TSVF). Here too, physical variables of the quantum system are determined by pre- and postselection, namely the initial measurements (preparation) and the final one. The two state vectors proceeding from these boundary conditions to the future and the past, respectively, give a full account of the evolution that took place between them. Moreover, striking phenomena like a particle disappearing and reappearing between distant parts of the wave function have been rigorously predicted by this formalism \cite{Disappearing,Spooky,WR}, awaiting experimental validation similar to \cite{Shutter2}.

Of special interest are the ``odd'' physical values derived by TSVF, which, due to the equivalence with standard quantum theory, are obliged by the latter as well. These are momentary values, prevailing between special pairs of pre- and postselections. A particle's mass, for example, can be extremely large, small or even negative. Momentary pairs of particles and nega-particles springing from the particle prior to final detection can cancel one another, leading to its disappearance, then part again, leading to its reappearance, as in \cite{Disappearing}.

Let us now apply this method to our crossing MZIs experiment (Sec. \ref{crossing}). The explanation is natural: The interaction is finalized by the two detectors A and B which detected the molecule and the electron by emitting the backward-evolving wave function. Here again, the two opposite waves (or weak values with opposite signs in the TSVF)can give rise also to {\it destructive} interference. Hence, each of the ``no wave'' segments which have puzzled us is the result of two opposite waves which cancelled each other. Bearing this in mind, we have a simple answer to the question: Why do we need to keep free also the paths where no wave was supposed to traverse? This avoidance is required in order to let the wave from the future proceed from the detector back to the corpuscle, cancelling the wave coming from the past source. It is this combination of forward- and backward-evolving waves that determine in retrospect the positions of the electron and molecule thus granting them real positions.

Note that this is a fully realist description.  It also preserves causality, so long as we accept causal histories which move back and forth in terms of the overall global time coordinate.

The situation is slightly more complex with the particle appearing and disappearing along different spacetime trajectories like in Sec. \ref{shutter}. Let us recall that we have  picked  an unusual  pair  of pre-  and  postselections.  These  are past  and  future
boundary conditions that,  when  naively  followed forward and  backwards, would normally give different, almost  conflicting histories.  And  yet, together, they  give the accurate history of this  surprising evolution (see \cite{Disappearing} for the  full mathematical analysis in terms  of weak values  and \cite{Spooky} for the detailed description of the shutter-probe interaction). In such cases, Nature seems  to ``go out of its way'' to fit together the unusual pair of two evolutions, imposed from both temporal ends.  We have, in other  words, different histories  that  nevertheless share an origin  (the splitting of the particle into three boxes) and  destination (the particle's reunification). These shared points  give rise to spacetime zigzags  through which  matter and energy can be exchanged between the remote boxes. The final outcome strikes  us as non-intuitive only because  we assign  it one evolution where there  are actually been two or more thereof, ``revising'' one another. Indeed, here too, despite the evolution's oddity, the same  condition holds  as in the previous example:  All possible spacetime trajectories allowed by the wave functions involved must remain open, even where nothing seems to go through them; any obstruction would ruin the results. In terms of our model, a spacetime region where ``nothing'' seems to have happened is rather the result of events  happening, followed by events  which  reverse their action.  Hence the causal efficacy of events  that could have occurred even if they never  have, as in IFM \cite{IFM}.

The addition of the backward-evolving wave function (sometimes perceived as a hidden variable residing in the future of the system \cite{Future}) enriches the description of quantum phenomena by providing information regarding incompatible variables. For example, a wave function prepared with a negligible position uncertainty can later be measured very accurately for its momentum. According to the retrocausal account, during intermediate times the effective description of the system consists of accurate position {\it and} momentum, outsmarting in a sense the uncertainty principle. This effective description can be experimentally probed when creating a weak enough coupling to the pre- and postselected system \cite{PP1} and then it results in the so-called ``weak value'' \cite{AAV} according to the TSVF approach. Interestingly, weak values have recently been inferred also with some standard strong measurement schemes \cite{Shutter1,Shutter2,Disappearing,Spooky,CP}.

This interpretation, which except from the addition of retrocausal effects is quite simple and realistic,  indicates that  something very  profound about  time's nature is still ill-understood. But have we not suspected it all along?

In the companion paper \cite{RETC} we describe one approach to how a form of retrocausality may emerge in a theory which is fundamentally both causal and irreducible.


\section{Conclusion:  The Price of Realism}

In this paper we have considered three thought experiments that appear to present a challenge
to naive  realism underlying the dBB pilot wave  theory.  We showed that in each case a coarse grained version of dBB can answer the challenge by correctly  accounting for the mysterious behavior.

This has admittedly come with a price, namely the fourth assumption {\bf (D)} of Sec. 2.1, which enforced the conservation of momentum and the principle of inertia, particularly requiring that particles interact with particles. This cannot be true.  Rather there are necessarily processes in which it appears that a particle bounces off an empty or ghost wave function, and conversely a particle may be as penetrable as a wave. This behaviour does account for the described phenomena, but in a way that appears to undermine the distinction between waves and particles, based on naive ideas of classical physics.

These subtleties in pilot wave theory are, admittedly, not new; they were understood to some degree by de Broglie \cite{nosurprise}, and are no surprise to contemporary experts in pilot wave theory \cite{experts}.  Moreover the fact that the guided ``particles'' do not obey the principle of inertia and conservation of momentum was understood by both de Broglie and Einstein, and may have been the basis for the latter's rejection of pilot wave theory as a candidate for the realist completion of quantum physics he sought.

We then raised the possibility that a retrocausal extension of pilot wave theory might offer a realist resolution of the puzzles of quantum theory that does less damage to our intuitive ideas of waves and particles.
This is a modification of de Broglie and Bohm's theory in which the guidance equation is a wave function with two components, one of which is the de Broglie guidance wave, and the other a copy of it which moves from the future into the past.  The sum then acts as the guidance wave, which moves the particle via the de Broglie guidance equation.

Continuing the line of inquiry proposed here, a second paper \cite{RETC} addresses the question: Do energetic causal sets models violate causality in a way similar to the retrocausality discussed in this work? This was indeed the initial challenge bringing the authors together and which sparked this collaboration.

\section*{Acknowledgements}

We are grateful to Yakir Aharonov, Andrew Liddle and Antony Valentini for many helpful discussions. We also wish to thank an anonymous referee for many helpful comments.

This research was supported in part by Perimeter Institute for Theoretical Physics. Research at Perimeter Institute is supported by the Government of Canada through Industry Canada and by the Province of Ontario through the Ministry of Research and Innovation. This research was also partly supported by grants from NSERC and FQXi. M.C.\ was supported by Funda\c{c}\~{a}o para a Ci\^{e}ncia e a Tecnologia (FCT) through grant SFRH/BPD/111010/2015 (Portugal). LS and MC are especially thankful to the John Templeton Foundation for their generous support of this project. Further, this work was also supported by Funda\c{c}\~{a}o para a Ci\^{e}ncia e a Tecnologia (FCT) through the research grant UID/FIS/04434/2013.

\end{document}